# Charging and coagulation of dust in protoplanetary plasma environments

L. S. Matthews, V. Land and T. W. Hyde
*Center for Astrophysics, Space Physics, and Engineering Research, Baylor University, Waco, TX, USA, 76798-7316*

Lorin_matthews@baylor.edu

ABSTRACT

Combining a particle-particle, particle-cluster and cluster-cluster agglomeration model with an aggregate charging model, the coagulation and charging of dust particles in various plasma environments relevant for proto-planetary disks have been investigated. The results show that charged aggregates tend to grow by adding small particles and clusters to larger particles and clusters, leading to greater sizes and masses as compared to neutral aggregates, for the same number of monomers in the aggregate. In addition, aggregates coagulating in a Lorentzian plasma (containing a larger fraction of high-energy plasma particles) are more massive and larger than aggregates coagulating in a Maxwellian plasma, for the same plasma densities and characteristic temperature. Comparisons of the grain structure, utilizing the compactness factor, $\varphi_\sigma$, demonstrate that a Lorentzian plasma environment results in fluffier aggregates, with small $\varphi_\sigma$, which exhibit a narrow compactness factor distribution. Neutral aggregates are more compact, with larger $\varphi_\sigma$, and exhibit a larger variation in fluffiness. Measurement of the compactness factor of large populations of aggregates is shown to provide information on the disk parameters that were present during aggregation.

*Subject headings:* accretion disk — dust — planets and satellites: formation — plasmas — protoplanetary disks

## 1. The formation of planets

At the time of writing more than 500 exoplanets have been observed with more than 400 of these confirmed, and more planets are being detected and confirmed on a weekly basis[1]. Even though these discoveries show that the process of planet formation is in itself a general one, they have also shown that our Solar System is everything but the perfect example of the average planetary system, Pluto, or no Pluto.
Partly due to the inherent bias of the available observational techniques, many of the earliest discovered systems involved large gaseous planets orbiting close to the parent star and planets on very eccentric orbits, much in contrast with our Solar System (Ollivier et al. 2009). Since many early planet formation theories were based on the Solar System (and in many cases these were then tested against our Solar System), these observations make clear that our knowledge of planet formation is incomplete.

The environment in which planet formation takes place is generally accepted to be a proto-planetary disk (PPD), a disk of gas and small (nanometer to millimeter sized) dust particulates accreting matter onto the central young stellar object (YSO), a famous example of which is observed around β-pictoris (Smith & Terrile 1984). There are many different models for the formation of gaseous and rocky planets in such disks: the *core-accretion* model (Pollack 1984), accompanied by *planet migration* theories (Raymond et al. 2006), or *disk-instability* models (Boss 1997). All of these involve an early stage in which the dust particles collide and stick together to form the initial seeds for planetesimal formation. This earliest stage is the topic of this paper, a stage which sets the starting point and initial conditions for the various planet formation theories and models currently under debate.

---

[1] http://www.exoplanets.org, http://www.exoplanet.eu

One of the oldest problems in accretion disk studies is the apparent inability of the gas (due to the minute viscosity of the gas) to transfer angular momentum outwards, while transferring mass inwards. A possible solution to this problem was presented in the α-*model*, in which turbulent viscosity provides sufficient friction for momentum transfer (Shakura & Sunyaev 1973). However, no specific mechanism providing the turbulence was presented; the presence thereof was simply assumed.

Two decades later, a connection was made between a magnetic instability in Couette flows, called the magneto-rotational instability (MRI) (Velikhov 1959), and the turbulence arising in accretion disks. This has become the most popular theory for the description of turbulence in accretion disks (Balbus & Hawley 1991). With this theory, however, arises one of the current dichotomies in studies of the micro-physics of the early stages of planet formation.

The MRI present in accretion disks *requires* the presence of an (albeit weakly) ionized medium, since the magnetic field involved has to be (albeit weakly) coupled to the disk matter. On the other hand, almost all studies to date involving the collision and subsequent sticking of microparticles have considered only neutral particulates in a gaseous environment. This is true for both theoretical and numerical studies (Kempf et al.1999; Dominik & Tielens 1997; Zsom et al. 2011), as well as experimental studies (Blum et al. 2000), although a few experiments examining magnetic grains have been performed (N¨ubold et al. 2002). Charging effects are usually considered to be insignificant, or are merely mentioned as an after-thought (Blum & Wurm 2008). Recently, simulations have shown that the effect of even a very modest grain charge can not be neglected; as such, a self-consistent charging/coagulation approach for plasma environments relevant to PPDs seems to now be in order (Matthews et al. 2007; Okuzumi 2009). Recent experiments have also provided evidence for run-away growth induced by electrostatic dipole interactions, showing that charging of particles can in fact speed up the coagulation processes (Konopka et al. 2005), as was also shown numerically (Matthews & Hyde 2009).

In this paper we present a numerical study of the coagulation and charging of dust particles and aggregates in different plasma environments relevant to PPDs. Section 2 discusses the different approaches used, while in section 3 the initial conditions for the simulations are discussed. Section 4 includes our results, while section 5 concludes with a discussion.

## 2. Charging and collisions of aggregates

The results in this study were obtained using a Particle-Particle/Particle-Cluster/Cluster Cluster Agglomeration (PPA/PCA/CCA) model, *Aggregate Builder*, coupled to an aggregate charging code, *Orbital Motion Limited Line Of Sight* (OML LOS). This section will briefly explain OML theory, as well as the line of sight approximation used to calculate the charge on the aggregates. Next, the principles of Aggregate Builder will be discussed. Since the fluffiness of particles is an important property, we will also briefly explain two ways of defining this parameter, namely through the *fractal dimension* and the *compactness factor*.

### 2.1. OML theory and Line Of Sight approximation

There are many mechanisms that result in the charging of dust in a plasma environment, including photodetachment (for instance by UV radiation), secondary electron emission (through the impact of energetic electrons), radioactive charging, tribo-electric charging and more (Mendis & Rosenberg 1994). However, in this study, we limit ourselves to charging through the collection of charged particles from the surrounding plasma.

The charging of a single particle immersed in plasma is described by OML theory, originally derived for Langmuir probe measurements (Allen 1992). The current density due to incoming particle species α (here we assume α = e or α = + for electrons and ions carrying one positive electron charge, respectively) to a point on the surface of a particle is given by

$$J_\alpha(t) = n_\alpha q_\alpha \int \int \int f_\alpha(v_\alpha) v_\alpha \cos(\theta) d^3 \vec{v_\alpha}, \quad (1)$$

with $n_\alpha$ the plasma density very far from the particle, $q_\alpha$ the charge of the incoming plasma particle, $f_\alpha(v_\alpha)$ the velocity distribution function of the plasma particles and $v_\alpha \cos(\theta)$ the velocity component of the incoming plasma particle perpendicular to the surface. By using $d^3 v_\alpha = v_\alpha^2 dv_\alpha d\Omega$, with $d\Omega$ the solid angle extended at the surface, we can split the integral to obtain

$$J_\alpha(t) = n_\alpha q_\alpha \int_{v_m(t)}^{\infty} f_\alpha(v_\alpha) v_\alpha^3 dv_\alpha \int \cos(\theta) d\Omega = n_\alpha q_\alpha \int_{v_m(t)}^{\infty} f_\alpha(v_\alpha) v_\alpha^3 dv_\alpha \int_0^{\frac{\pi}{2}} \cos(\theta) \sin(\theta) d\theta \int_0^{2\pi} d\phi. \quad (2)$$

Here, $v_m(t) = 2\ (\ )$ is the minimum velocity a plasma particle having the same charge compared to the dust particle must have to reach the dust particle surface, with $V_D$ the surface potential. For plasma particles with the opposite charge, $v_m = 0$.

For a single spherical particle the integral over the solid angle is trivial, but for an aggregate the integral becomes complicated, since one monomer in the aggregate can block part of the solid angle available to a point on another monomer in the aggregate, as illustrated in Figure 1, on the left. As an example of this, the *open* solid angle (in this 2D representation) for four points on one of the monomers in a small aggregate is indicated by the dashed areas.

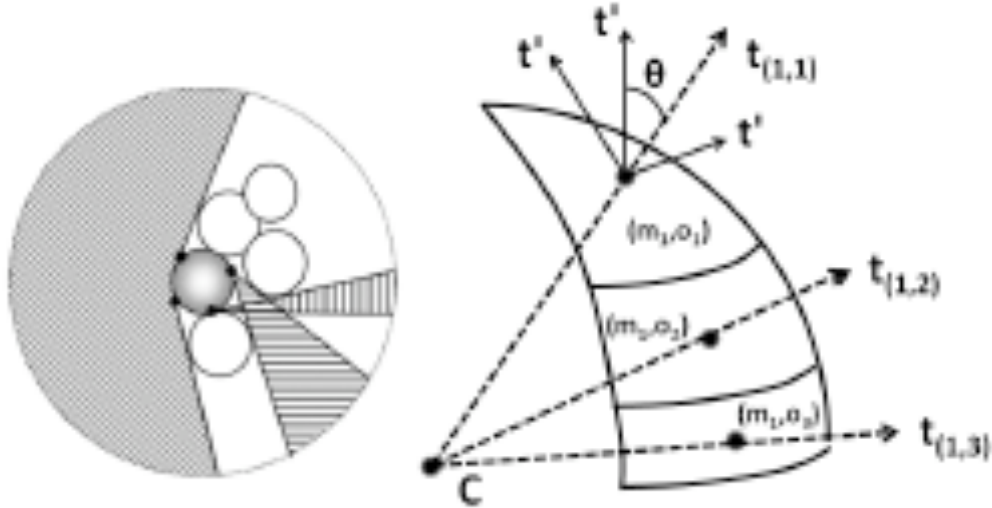

Fig. 1.— A 2D representation of the OML LOS geometry, as shown on the left. Different points on a monomer in an aggregate are partly shadowed from the outside plasma by other monomers. The dashed areas indicate the open, unblocked lines of sight for four points on a monomer. These unblocked lines of sight are used to approximate the solid angle in the integral for the current density. An illustration of the test directions defined in the model to calculate the current density to a surface patch on one of the monomers is shown on the right. C indicates the center of the monomer with three surface patches having coordinates (m1, o1), (m1, o2), (m1, o3) shown. The vectors **t**, which are also the normal vectors to the surface patches, show some of the test-directions for determination of the open lines of sight. The vectors **t** indicate such directions applied on one of the patches. When one of these vectors intersects another monomer in the aggregate, or points inwards into the monomer the patch sits on, a 0 is assigned to that direction, otherwise a 1. $\cos(\theta)$ is calculated for all test-directions to provide the components normal to the surface patch.

(A color version of this figure is available in the online journal.)

In order to calculate the current density of species $\alpha$ to a monomer in the aggregate at a given time, $J_\alpha(t)$, the surface of the monomer is divided into equal-area patches with coordinates $(m, o)$, similar to the longitude-latitude system, but with each surface patch having the same area, $A(m,o)$. Vectors pointing from the center of the monomer to these surface points define $[m \times o]$ test directions, $\mathbf{t}(m,o)$. At a point $(m, o)$, the test directions **t** originating from that point (the so called lines of sight), are determined to be *blocked* if they intersect any other monomer in the aggregate, or the monomer in question, and *open* otherwise. Each line of sight is then assigned a value of 0, or 1, respectively. The number of 1's divided by the total $[m \times o]$ number of test-directions comprises the *open lines of sight factor* for the patch at point $(m, o)$, $LOS(m,o)$. At the same time, the cosine of the angles between the normal direction of the patch and each

**t** is also determined and then taken into account for LOS(m,o). Figure 1 illustrates this method, on the right.

The net current of species α to a patch at point (m, o) at a given time, I$_α$,(m,o)(t), can now be found by multiplying the current density by the area of the patch times the open line of sight factor for the patch (with the cosine factors): I$_α$,(m,o)(t) = J$_α$(m,o)(t) × A(m,o) × LOS(m,o). Summing over the species α provides the change in the surface charge on the patch during a time interval dt, dQ$_D$,(m,o)(t) = ΣI$_α$,(m,o)(t) dt. Note that the current to the patch depends on v$_m$(t), which in turn depends on V$_D$,(m,o)(t) (hence on Q$_D$,(m,o)(t)), so that the solution requires numerical iteration until equilibrium is reached. The change in charge of the monomer is then obtained by adding up the contribution of all the patches. The change in charge of the aggregate as a whole is obtained by adding the contribution from each of the N monomers. This process is iterated in time until the change in aggregate charge becomes negligible, dQagg < 0.0001%, at which point on average the net current to the aggregate will be zero.

For sufficient resolution, we use over 400 patches per monomer (m×o = 20×21). The largest aggregates in this study contain just over 2000 monomers. Since each time iteration requires on the order of 100 time steps, the total charge calculation for the largest aggregates require on the order of $10^8$ iterations. Obviously, the computations for determining the detailed charge structure are very time consuming.

The dipole moment for the aggregate is found in a similar manner. The contribution of each patch is computed and then summed to obtain the dipole moment on each monomer. To obtain the dipole moment of the aggregate as a whole, the contribution of each monomer is then added up. The monopole and dipole charges are then used in the calculations of the collisions performed in Aggregate Builder, as discussed below.

## 2.2. The plasma environment: $f_α(v_α)$

In this paper, we will consider three different environments: neutral gas, a Maxwellian plasma, and a Lorentzian plasma, which is often observed in space plasma environments (Scudder 1994). For neutral gas, the OML LOS routine is not used and the charge on the aggregates is set to 0. In a plasma environment where local thermal equilibrium holds (and collisions are important), the Maxwellian distribution is used:

$$f_\alpha(v_\alpha) = \left(\frac{m_\alpha}{2\pi k_B T_\alpha}\right)^{3/2} \exp\left(-\frac{m_\alpha v_\alpha^2}{2 k_B T_\alpha} - \frac{q_\alpha V_D}{k_B T_\alpha}\right), \tag{3}$$

with $V_D$ the dust particle surface potential and $m_\alpha$ and $T_\alpha$ the electron/ion mass and temperature, respectively. In a plasma environment where momentum transfer collisions are less frequent, the plasma populations acquire a larger high velocity tail in the distribution, leading to a so-called Lorentzian, or κ-distribution,

$$f_\alpha(v_\alpha) = (\pi\kappa\theta^2)^{-3/2}\frac{\Gamma(\kappa+1)}{\Gamma(\kappa-\frac{1}{2})}\left(1 + \frac{v_\alpha^2 + q_\alpha V_D}{\kappa\theta^2}\right)^{-(\kappa+1)}. \tag{4}$$

Here, Γ(x) is the gamma function and $\theta = v_{T,\alpha}[(2\kappa - 3)/\kappa]^{1/2}$ is a generalized thermal speed, with $v_{T,\alpha} = $ . The distribution is defined for $\kappa \geq 3/2$ and tends to the Maxwellian distribution for $k \to \infty$. For typical astrophysical environments, $\kappa = 5$ is generally assumed (Mendis & Rosenberg 1994), which will be done here as well.

For an isolated spherical particle in a Maxwellian hydrogen plasma at thermal equilibrium, the particle potential becomes $\varphi_D \approx -2.51 k_B T/e$ (T in K), whereas for a Lorentzian plasma the surface potential (and therefore the charge) will become more negative with decreasing κ.

## 2.3. Aggregate Builder

Aggregate Builder is based on an N-body code originally developed to investigate the gravitational interactions between planetesimals and objects in rocky rings (Richardson 1993). The code has since been modified and extended to include the effects of charged particles and magnetic fields (Matthews & Hyde 2003; Vasut & Hyde 2001; Qiao et al. 2007). The modified code treats accelerations caused by interactions of charged grains as well as rotations induced by torques due to the charge dipole moments (Matthews et al. 2007). These dipole-dipole interactions have been shown to greatly enhance the collision rate, even for like-charged particles (Matthews & Hyde 2009).

Aggregate Builder is used to study pairwise interactions of colliding particles in the COM-frame of the target particle. The incoming particle has a velocity directed towards the target particle to within an offset-distance $(a_t + a_i)/2$ of the COM, where $a_t$ and $a_i$ are the maximum radii of the target and incoming particles, respectively, as indicated in figure 3. Libraries of aggregates are created from successful collisions, which can then be used as starting points in N-body codes. This allows modeling of the aggregation of large distributions of aggregates. Information on the missed collisions is saved for collision statistics.

Aggregates are built in three steps. First generation aggregates are built by additions of single monomers, up to a size of N = 20. A charged monomer (with a charge determined by the plasma temperature and monomer size) is randomly chosen from a size distribution and placed at the origin. Another randomly selected monomer is "shot" towards the first monomer. The initial velocity is a combination of Brownian motion and the velocity derived from turbulence theory, and as discussed in section 2.4, these velocities are shown to be low enough for grains to stick at the point of contact without fragmentation. The orientation and position of each particle is tracked and a collision is detected only when two monomers actually overlap. The new aggregate properties are then updated, including the charge (calculated with OML LOS), and the resultant aggregate is saved to the first generation aggregate library. The origin is then defined to be the center of mass (COM) of the new aggregate and the process is repeated.

The second generation is constructed by randomly selecting one of the aggregates from the first generation library, containing aggregates with 2 ≤ N ≤ 20, and placing it at the origin. In 60% of the cases a randomly selected monomer is shot towards the aggregate, while in 40% of the cases another aggregate randomly chosen from the library is used. Both the monopole and dipole interactions can induce torques on the aggregates, causing them to spin and their trajectories to deviate from straight lines. Figure 2 shows the importance of such dipole interactions. When dipole interactions are ignored in the calculations, the resulting aggregate has an entirely different geometry and fluffiness. Aggregates containing up to at least 200 monomers are constructed this way, charged in OML LOS and stored in the library for second generation aggregates.

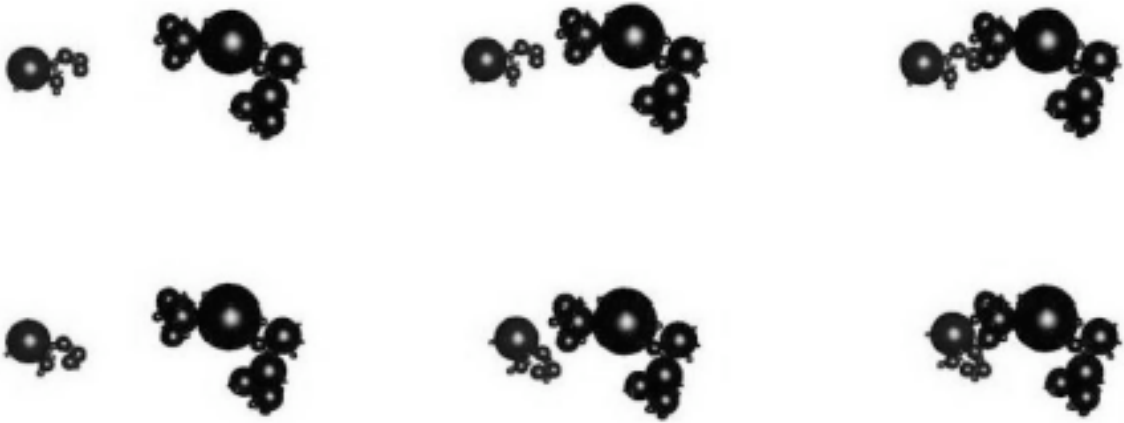

Fig. 2.— An example showing the importance of dipole interactions. The top row shows three snapshots of a collision between two charged aggregates when dipole interactions are ignored. The particles approach along a straight-line trajectory, as in a ballistic collision, with the particles slowing as they approach each other. The bottom row shows another collision between the same aggregates, but now the dipole interactions are taken into account. The rotation of the aggregates results in a completely different geometry of the final aggregate. In this case, the interaction time between the two aggregates is also longer due to the rotation changing the orientation of the aggregate.

(A color version of this figure is available in the online journal.)

Finally, the third generation is created by randomly selecting an aggregate from the second generation library, containing aggregates with 3 ≤ N ≤ 200, and placing it at the origin. In 40% of the cases the incoming particle is a monomer, in 30% an aggregate from the first generation library and in the remaining 30% an aggregate from the second generation library, resulting in the growth of aggregates with N ≈ 2000. These aggregates are then charged in OML LOS.

2.4. Relative particle velocities in turbulent flows

Assuming a capacitor model for dust particles, the charge can be related to the surface potential by $Q_D = 4\pi\varepsilon_0 V_D a$, with a the particle radius. For a Maxwellian hydrogen plasma this gives $Q_D \approx -10\pi\ \varepsilon_0 k_B T a/e$, or in more useful units, $Q_D \approx -1700e \times T(eV) \times a(\mu m)$. For two particles having radii $a_1$ and $a_2$ and charge $Q_1$ and $Q_2$ to collide and stick, they must have sufficient energy to overcome the repulsive Coulomb interaction between them.

Their initial kinetic energy, $K = 0.5\mu v_r^2$, with $\mu = m_1 m_2/(m_1 + m_2)$ the reduced mass and $v_r$ the relative particle velocity, should therefore be larger than $U = Q_1 Q_2/4\pi\varepsilon_0 \Delta$ with $\Delta = a_1 + a_2$. The minimum relative velocity is then given by $v_{r,min} = 2$ .

As an example, the repulsive interaction energy between a 0.5 μm and a 10 μm silicate particle (with mass density $\rho$ = 2.5 g cm$^{-3}$) in our simulation in a hot PPD environment with T ≈ 0.1 eV, is given by $U = 85 * 1700 e^2/4\pi\,\varepsilon_0\,(10.5 \times 10^{-6}) \approx 20$ eV. This gives $v_{r,min} \approx 7$ cm s$^{-1}$. Brownian motion results in a relative particle velocity of $v_{r,B} = 8$ ≈ 0.5 cm s$^{-1}$, which is clearly not large enough to allow successful collisions between these charged particles. (Interestingly enough, the required velocity for two particles with a = 0.5μm is also 7 cm s$^{-1}$, which shows that Brownian motion is also unable to account for successful collisions between the smallest charged monomers in our distribution, with the smallest repulsive interaction energy.)

Even though there are different mechanisms that can provide additional velocities to a particle in a PPD, for example radial drift and gravitational settling (Brauer et al. 2008), it has been shown that the corresponding velocities are relatively small. In Rice et al. (2004) the radial drift for micron sized particles in a disk at 1 AU was found to be much less than 1 mm s$^{-1}$, while in Dullenmond & Dominik (2004) gravitational settling was found to result in velocities several orders of magnitude less than the local Kepler time. We therefore assume that in this case turbulence is the primary contributor to the relative velocities for dust particles in the PPD.

A complete derivation of the relative particle velocities in turbulent flow for different regimes is provided in Ormel & Cuzzi (2007). We here limit ourselves to the *small particle* regime, where dust is *strongly coupled* to the turbulent eddies. In this case, the relative velocity between two particles (indicated by subscripts 1 and 2) is given by

$$v_{turb} = \sqrt{\frac{3}{2}} V_\eta \frac{(t_1 - t_2)}{t_\eta}, \tag{5}$$

with $t_s = 3m/4 c_g \rho_g \sigma$ the stopping time of particle s, where $c_g$ and $\rho_g$ are the sound speed in the surrounding medium and the mass density of the surrounding medium, respectively, m is the dust particle mass and σ its geometric cross section (Weidenschilling 1984). $V_\eta$ and $t_\eta$ are the characteristic turn-over velocity and time scale, respectively, which are related to the velocity and time scale of the largest eddies, on the Lagrangian scale, through the Reynolds number of the turbulent flow: $V_\eta \sim L_\eta/t_\eta$, $L_\eta = Re^{-3/4} L_L$ and $t_\eta = t_L Re^{-1/2}$.

At a distance r from the YSO, the Lagrangian scales are typically chosen as the Keplerian orbit time, $t_L = t_K \sim (\Omega_K(r))^{-1}$ and $L_L$ = r. The Reynolds number in astrophysical situations is usually described by the α-model (Shakura & Sunyaev 1973), $Re = \alpha c_g^2/\nu\Omega$, with ν the viscosity of the gas in the disk and Ω the rotation frequency and α a numerical constant that is varied. In simplest terms, this model relates the thickness of the disk to the turbulence in the disk, since $c_g/\Omega \sim H$, the vertical scale height of the disk. Using the proper disk and turbulence parameters, the relative velocities are then computed.

## 2.5. Defining fluffiness: Fractal dimension and compactness factor

The fluffiness of an aggregate is very important to the coagulation process, since a fluffier aggregate couples more effectively to the gas in PPDs than does a compact aggregate (Nakamura & Hidaka 1998). It also has different optical properties than a compact particle (Hage & Greenberg 1990), can provide a larger surface area for chemical reactions occurring in PPDs (Ehrenfreund 2003) and has a larger collisional cross-section. In order to provide consistent statements about the fluffiness of the aggregates resulting from our simulations, we therefore need a consistent measurement, which is computationally robust and inexpensive at the same time.

One method commonly used is the fractal dimension. The method employed in our simulations is the Hausdorff dimension. In this method we place a box around the aggregate, with side $a_0$, so that the largest dimension just fits inside the box, and then divide this box into small sub-boxes with size $a \ll a_0$. We then count the number of sub boxes, $N$, that are occupied by any part of the aggregate to obtain the fractal dimension as

$$d_F = \frac{\log N}{\log(a_0/a)}. \tag{6}$$

In the limit of a completely compact aggregate the above approaches three, and in the limit of a completely linear aggregate it approaches one. Although this represents a relatively simple method, the outcome depends strongly on the size of the sub-boxes used (hence the number of boxes to count) and becomes computationally rather expensive. Furthermore, the side length of the sub-boxes, $a$, has to be larger than the diameter of the monomer, which makes this method difficult to apply to polydisperse monomer distributions, such as those being considered in this study.

A more contemporary method is based on the idea of a compactness factor (Min et al. 2006). In this case, the aggregate is projected on a plane and the projected surface area $A_i$ is calculated. The equivalent radius, $R_i$, is then defined by equating this area to the area of a circle, $A_i = \pi R^2_i$. Averaging this over many orientations yields an average equivalent radius, $R_\sigma$. The ratio of the total volume of all the monomers to the volume of the equivalent sphere with radius $R_\sigma$ then defines the compactness factor

$$\phi = \frac{\sum_{j=1}^{N} r_j^3}{R_\sigma^3}, \tag{7}$$

with N the number of monomers in an aggregate. For fluffy aggregates, φ approaches 0, for compact aggregates it approaches 1. An illustration for a representative aggregate is shown in Figure 3. Even though this method is more complex, it provides a much more robust determination of the fluffiness of aggregates and is less expensive computationally.

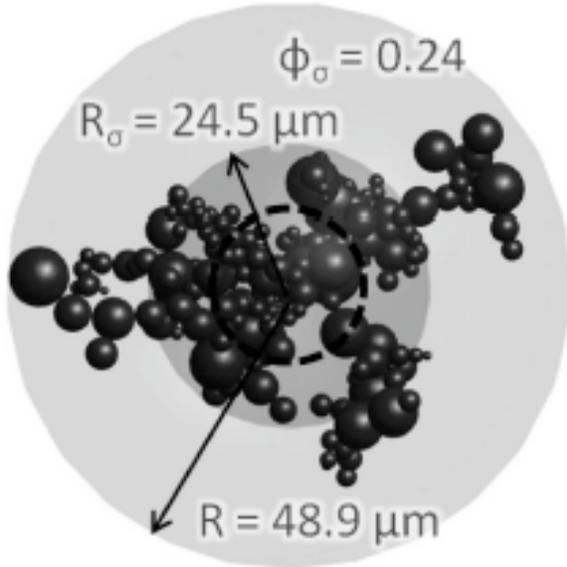

Fig. 3.— An illustration of the compactness factor for one aggregate. The lighter colored circle indicates the maximum aggregate radius, R, defined as the maximum extent of the aggregate from the COM, which for a single monomer would equal the monomer radius, a. The darker inner circle indicates $R_\sigma$, as defined in the text. For this aggregate the compactness factor, $\varphi_\sigma$, is calculated as 0.24. This means that the total volume of all the monomers fills a smaller equivalent sphere of radius 0.68 $R_\sigma$, as indicated by the black dashed circle.

(A color version of this figure is available in the online journal.)

## 3. Initial conditions for the simulations

Initially, monomers are assumed to be silicates with radii in the range 0.5 μm < $a_0$ < 10 μm, distributed according to the MRN distribution, $n(a_0) \propto a_0^{-3.5} da_0$ (Mathis et al. 1977), which gives an average monomer radius of $<a_0> = 0.82$ μm. This is typical for grains in PPDs, although the size distribution is probably less steep, see Weidenschilling (1984). Silicates are commonly observed in promordial material in the Solar System and specific absorption features for silicates are routinely observed in spectra from PPDs. Furthermore, silicates are non-conducting particles, for which charge can be assumed to stick at the point of impact, exactly as our model assumes for aggregate charging.

The PPD is assumed to have a 1 M☉ central mass, mass density of $\rho_g = 2 \times 10^{-6}$ kg m$^{-3}$ and mid-plane temperature of roughly $T_{g,mp} \sim 1200$ K (Ruden & Pollack 1991; Lissauer 1993). Using the typical value of α = 0.01, with the viscosity defined by $v(z) = 9 \cdot 10^{-10} c_g(z)/\rho_g(z)$ (Weidenschilling 1984), the Reynolds number is found to be Re = 2 · 10$^{12}$, 1 AU from the central mass and at z = H/2. We find $t_\eta$ = 21 s. Since we require $t_{s,max} < t_\eta$ for strongly coupled particles, the largest particles that would still be coupled for these conditions have radii of 12 cm, much bigger than our largest size of 10 μm. Similarly, we find $L_\eta$ = 89 m and $V_\eta \sim L_\eta/t_\eta$ = 4.2 m s$^{-1}$.

The maximum turbulent relative particle velocity between monomers will occur for a monomer of 10 μm radius and one of 0.5 μm radius, for which we find $(t_1 - t_2) \sim 2$ s. For atomic hydrogen at 900 K $c_g$ = 2900 m s$^{-1}$. The maximum turbulent relative velocity becomes $v_{turb,max}$ = *3/2* ∗ 4.2 ∗ 2/21 = 0.49 m s$^{-1}$, an order of magnitude larger than $v_m$. This velocity is near the restructuring threshold, but much lower than the velocity needed for destruction (Wurm & Blum 1998; Blum & Wurm 2000).

In order for restructuring to play a significant role for the morphology of aggregates, the kinetic energy requirement is given by K > 5$E_{roll}$, where $E_{roll} = 6\pi^2 \xi \gamma \alpha_\mu$ is the restructuring energy (Ormel et al. 2009). $\alpha_\mu = a_1 a_2/(a_1 + a_2)$ is the reduced radius for two colliding aggregates, γ is the surface energy density for the material and ξ is the critical displacement for a monomer in an aggregate after which restructuring starts. In Blum & Wurm (2000) γ was measured for pure silicate, SiO2, as 0.019 J m$^{-2}$, while in Ormel et al. (2009) the value for *ices* was reported as 0.37 J m$^{-2}$, while ξ was taken as 2 nm. For our example of the two monomers above, this results in a restructuring energy of $E_{roll}$ = 6.7 keV. Their kinetic energy is only 0.98 keV, however.

The real question of course is whether or not a collision between aggregates, or between monomers and aggregates can lead to restructuring. Since the kinetic energy depends quadratically on the velocity, the relative turbulent velocity will be the most important factor. The largest relative velocity occurs between the smallest monomer and the largest aggregate. The largest aggregate has R ≈ 100 μm with a mass of roughly $10^5 <$ m0 $>\approx 6.2 \times 10^{-10}$ kg. Estimating the cross section σ ≈ π$R^2$ ≈ 3.14 × 10$^{-8}$ m$^2$, we find a stopping time of $t_{agg}$ ≈ 2.6 s. The stopping time for the 0.5 micron monomer is 0.22 s. The relative turbulent velocity between the monomer and the aggregate then equals $v_{turb}$ = 0.58 m s$^{-1}$, so that the kinetic energy becomes K = 1.58 keV. The restructuring energy is Eroll = 7 keV, hence restructuring does not play a role here. Apparently, smaller particles that can couple to faster turbulence, or larger size differences than obtained in the simulations presented here, are required for restructuring to become important for aggregate morphologies.

## 4. Results

### 4.1. Collision statistics

We start by showing the collision statistics, since these contain much of the important information about the collision processes for the different plasma environments. Figure 4(a) shows the number of missed collisions per successful collision, while figure 4(b) shows the size of the incoming aggregates that successfully collided, versus the size of the resulting aggregate, in number of monomers per aggregate.

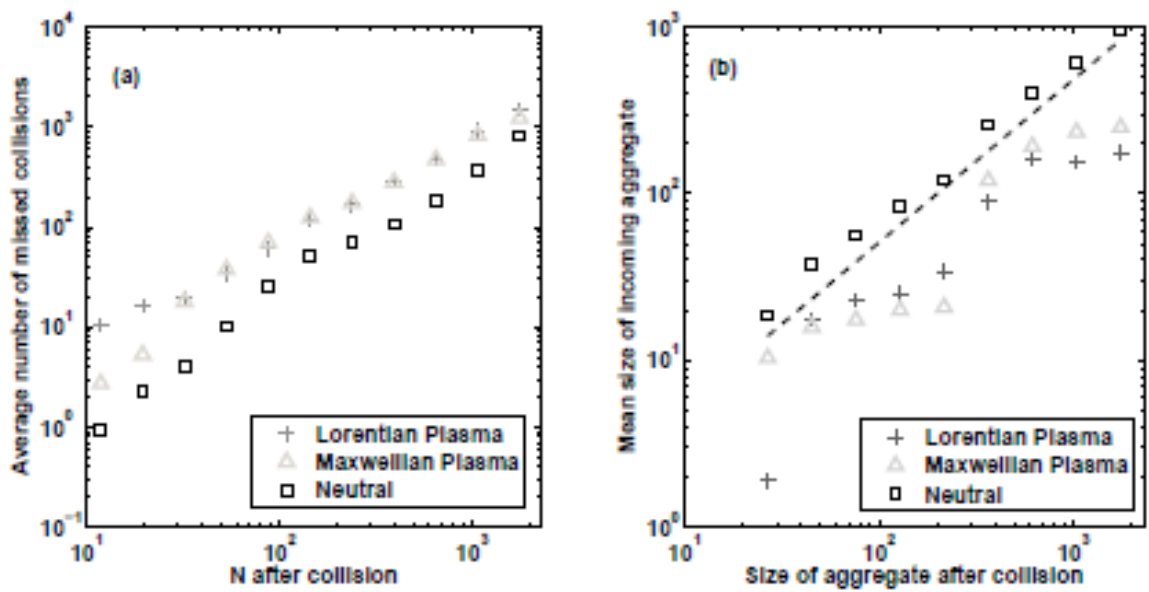

Fig. 4.— Collision statistics of aggregate collisions for three different plasma environments. The data for all the aggregates formed have been binned and averaged. (a) The number of missed collisions per successful collision against the number of monomers in the resulting aggregate. The squares indicate neutral aggregates, triangles and crosses charged aggregates in a Maxwellian and Lorentzian plasma environment, respectively. (b) The number of monomers in the incoming aggregate versus the number of monomers in the resulting aggregate. The neutral aggregates form through collisions of almost equally sized aggregates, which is indicated by the dashed line. The squares indicate neutral aggregates, the triangles and crosses charged aggregates in a Maxwellian and Lorentzian plasma environment, respectively.

(A color version of this figure is available in the online journal.)

Neutral aggregates exhibit the highest collision probability, at approximately 1/2 for the smallest aggregates to 1/1000 for the largest aggregates. The collision probability is smaller for charged aggregates in both plasma environments. Any discrepancy between the two plasma environments is only apparent for smaller aggregates, which shows that increased negative charge decreases their collision probability, while for larger aggregates, the collision probability is roughly the same for both plasma environments. This is due to the fact that even though aggregates in a Lorentzian plasma have a larger negative charge, they also are fluffier, as is shown below, which helps to increase the collision probability, since they have a larger cross section for the same number of monomers. Overall, the collision probability for the smallest aggregates is 1/3 (1/10) for the Maxwellian (Lorentzian) plasma and for the largest aggregates about 1/1200 (1/1800).

Successful collisions for neutral particles occur on average between aggregates of equal size (in fact, on average the incoming aggregate contributes 54% of the monomers in the resulting aggregate), whereas charged aggregation occurs between aggregates of distinctly different sizes, since the incoming aggregate is always much smaller than the resulting aggregate. This is to be expected, since charged particles need a larger relative velocity due to turbulence, which requires a large size difference between the colliding aggregates, as is clearly shown in figure 4 (b).
The jump visible between the second and the third generation comes from the sudden availability of larger aggregates for the incoming aggregate (picked from the 2$^{nd}$ generation library) and the strong likelihood that these participate in successful collisions, since they immediately introduce a large size difference. Note that the neutral aggregates do not exhibit such a jump. This jump for charged aggregation is somewhat artificial, since in a real dust cloud, the depletion of smaller aggregates will be a gradual process and not such a sudden imposed condition. Nonetheless, the aggregation of larger charged aggregates shows similar behavior, indicating that large size-differences enhance aggregation between charged aggregates.

As mentioned, aggregates formed in a Lorentzian plasma are fluffier than those formed in a Maxwellian plasma. This is shown in Figure 5. The compactness factor obtained for neutral aggregates falls roughly in between the compactness factor obtained for aggregates in the two plasma environments, however, the spread in the compactness factor is rather large. This is an indication of the absence of the electric multipole interactions, which clearly play an important role in the charged particle aggregation, either by allowing only extended *arms* of the aggregates to stick during collisions, or by alignment of the aggregates during successful collisions, resulting in fluffier, more porous aggregates, with a more peaked compactness factor distribution.

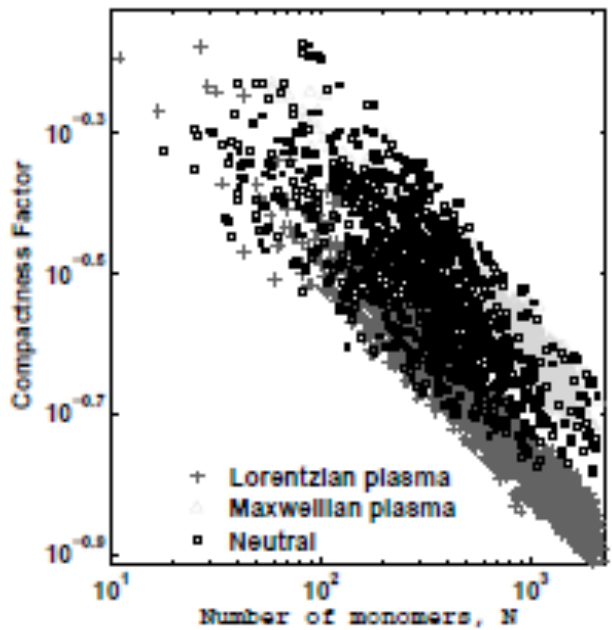

Fig. 5.— The compactness factor obtained for three different plasma environments (crosses correspond to aggregates in the Lorentzian plasma, triangles to aggregates in the Maxwellian plasma and squares to neutral aggregates), showing that the increased charge on aggregates formed in a Lorentzian plasma leads to much fluffier aggregates. Note that for clarity, aggregates in the first and second generation library are not included in this graph.

(A color version of this figure is available in the online journal.)

This is also shown in Figure 6 (a-c). For comparison, the fractal dimension is shown in Figure 6 (d-f), using the Hausdorff method. The compactness factor shows aggregates becoming fluffier with increasing size. Also, the compactness factor distribution becomes more peaked for larger charged aggregates, especially when compared to neutral aggregation. The distribution function for all aggregates taken together, including the aggregates in the first generation library, does not show these peaks clearly. This occurs because there are many more first generation library aggregates in the whole population than there are larger aggregates from the second and third generation libraries. However, the narrowing of the compactness factor distribution with increasing aggregate size for charged aggregates in part balances this effect. Therefore, for large aggregate populations, measurement of the compactness factor distribution should still allow plasma environments to be distinguished from neutral environments through the presence or absence of a small peak at low values of the compactness factor.

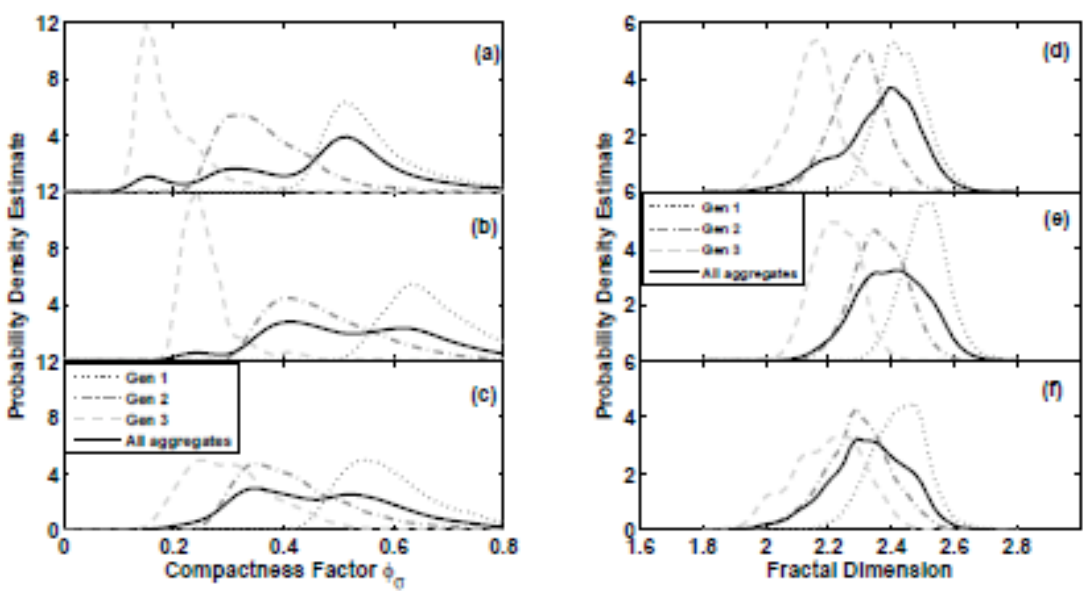

Fig. 6.— The probability distribution for the compactness factor (a-c) and the fractal dimension (d-f) for the three generations of aggregates, indicated by the different dashed lines, as well as all the aggregates together, indicated by the solid line, for the different plasma environments. From top to bottom the environments are a Lorentzian plasma, a Maxwellian plasma and a neutral environment.

(A color version of this figure is available in the online journal.)

These differences, when considered using the fractal dimension are far less pronounced and narrowing of the fractal dimension distribution for the later generations is much less visible. The overall distributions (when taking all generations together) have roughly the same median value ($D_f \approx 2.3$) and the contribution at lower values of $D_f$ for the charged aggregates is not clearly different from the contribution for neutral aggregates. Overall, it seems that the differences in fluffiness due to different plasma environments is more obvious when using an analysis based on the compactness factor than when using the fractal dimension (at least according to the Hausdorff dimension) and therefore distinguishing a plasma from a neutral environment should be more feasible using a measurement of the compactness factor of a large population then when using the fractal dimension.

Although the charge and dipole moments play an important role in charged particle aggregation, their calculation for aggregates quickly becomes computationally expensive. As such, for simulations of large aggregate populations, a heuristic fit for the charge and dipole moment based on some specific property of the aggregate would be valuable. Figure 7 shows fits for the charge and dipole moment obtained from our simulations for the Maxwellian and Lorentzian plasma environmnents. The fit for the charge depends on the number of monomers in the aggregate, whereas the dipole moment can be well fitted by using the maximum moment of inertia of the aggregates.

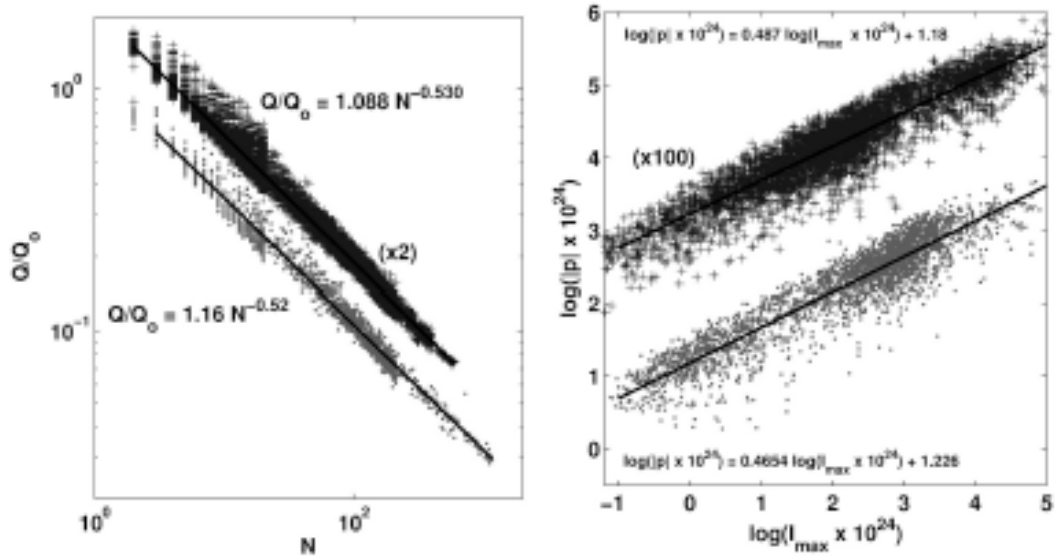

Fig. 7.— The charge normalized to the charge obtained through charge conservation versus the number of monomers in the aggregates for the Maxwellian plasma and Lorentzian plasma, and the magnitude of the dipole moment as a function of the maximum moment of inertia of the aggregates for the Maxwellian and Lorentzian plasma. Note the log-log scale. The black solid lines indicate linear fits to the log-log data.

(A color version of this figure is available in the online journal.)

The charge on an aggregate can be well approximated by the reciprocal of the square root of the number of monomers in the aggregate. This is useful when a complex N body calculation of large populations is needed (Matthews & Hyde 2004), since we do not have to separately charge all the aggregates using OML LOS, but instead can simply use the analytical fit, decreasing the required computational time significantly. A similar fit for the dipole moment has recently been published (Matthews & Hyde 2009), albeit with a large amount of scatter. Using the maximum moment of inertia, a more reasonable fit with far less scatter can now be obtained. Similar to the charge, this heuristic fit can now be used in large N-body simulations, rather than having to calculate it for every aggregate separately using OML LOS.

Finally, Figure 8 shows the mass, in units of average monomer mass, and the maximum radius of the aggregates, as defined in Figure 3, versus the number of monomers in the aggregate. The green points indicate neutral aggregates, red ones indicate charged particle coagulation in a Maxwellian environment and blue ones indicate a Lorentzian environment. Note that, again, we have excluded the data for the first generation, plotting only second and third generation aggregates.

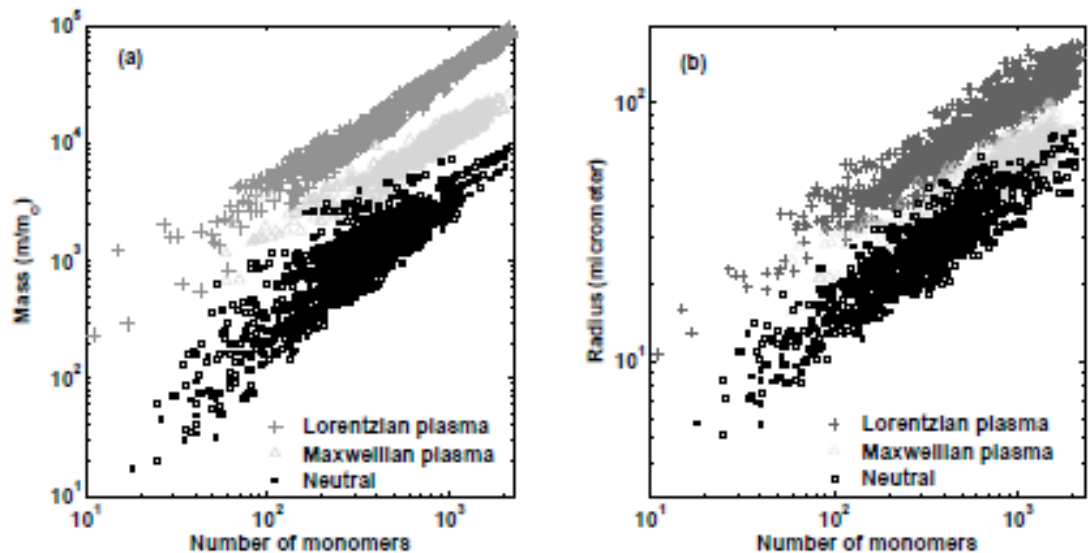

Fig. 8.— (a) The mass of the aggregates against the number of monomers in the aggregates. (b) The maximum radius against the number of monomers in the aggregates. The crosses indicate aggregates in the Lorentzian plasma environment, the triangles aggregates in the Maxwellian plasma and the squares indicate neutral aggregates.

(A color version of this figure is available in the online journal.)

It is clear that neutral aggregates are the lightest and smallest aggregates, the Maxwellian environment leads to heavier and larger aggregates, whereas the Lorentzian environment yields the most massive and largest aggregates of the three.

## 5. Discussion and conclusions

We have shown that the effect of charge on aggregation is very important for the formation of larger aggregates in PPDs. The main results can be summarized as follows:

• Charged aggregate coagulation leads to more massive, larger, and fluffier aggregates than does neutral aggregate coagulation.

• Collisionless (Lorentzian) plasma environments lead to more massive, larger, and fluffier aggregates than collisional (Maxwellian) plasma environments.

• An analysis of the fluffiness of aggregates using the compactness factor is more sensitive to smaller changes in the fluffiness than one employing the fractal dimension obtained using the Hausdorff dimension. Thus, distinguishing aggregation of a large population of aggregates in a plasma environment from aggregation in a neutral environment is more feasible when the compactness factor is used than when the fractal dimension is used.

- Collision statistics indicate that neutral particles collide more efficiently and that on average aggregates of roughly equal size collide. Charged particles collide much less efficiently and do so mainly through aggregation of smaller aggregates onto larger aggregates.

- A heuristic fit for the aggregate charge can be obtained in terms of the number of monomers in the aggregate, whereas the dipole moment can be obtained from the maximum moment of inertia of an aggregate.

Since momentum transfer collisions will most likely occur between charged particles and neutrals (due to the low ionization degree), charged particles are more likely to distribute according to a Lorentzian distribution in hot regions in PPDs where the ionization degree is slightly higher. The results for these distributions therefore most likely apply to the regions close to the central YSO and in the well-mixed regions outside of the *dead-zones*, even though recent simulations have shown that these dead-zones are likely to be much smaller than initially anticipated (Turner et al. 2007). Colder, less ionized regions, or regions ionized by external (cosmic and UV) radiation are more likely to develop Maxwellian distributions for the plasma, due to the lower ionization degree and higher collisionality. The results from our simulations assuming a Maxwellian distribution should therefore apply to regions in PPDs farther out from the central YSO, or for regions farther out still, but closer to the surface of the disk, where external radiation sources can cause additional ionization. Finally, the results for neutral aggregation apply best to the far outskirts of the PPDs, far away from the YSO, but closer to the midplane of the disk, where the gas is shielded from external radiation.

We therefore expect the biggest, most massive and fluffiest aggregates to predominantly form in the inner regions of the PPD, intermediately sized and more porous aggregates in Maxwellian environments farther out, and finally the lightest, smallest and most compact aggregates in the cold dusty outskirts of PPDs. Even though we started by stating that our Solar System is a bad example, as far as planetary systems go, it is still intruiging to speculate whether or not the distinction of the Ice Giants, Gas Giants and rocky planets in our Solar System might have anything to do with this.

The observed differences in the fluffiness of particles for the different plasma environments should also have an influence on the coupling of the aggregates to the surrounding medium, due to the differences in m/σ. The stopping times for fluffier aggregates will be shorter than the stopping times for compact aggregates for a given mass, which means that these aggregates can be more easily carried along by turbulent eddies with smaller turn-over times, $t_\eta$. Additionally, the increased surface area of fluffy aggregates will allow faster reaction rates for surface enhanced chemistry, believed to be important in PPDs, making our results important for large scale chemistry models of PPDs (Millar et al. 2003). Fluffy aggregates will also have different optical properties than compact aggregates (Min et al. 2006), so that the proper identification of aggregate distributions and their constituents, as well as the plasma environment in which they are immersed might well be derived from observations of the light originating from the dusty regions in PPDs.

Finally, if a direct relation between the optical properties of aggregates and their compactness factor can be obtained, either through laboratory measurement or numerical simulation, in-situ data on the compactness factor of aggregates collected in the outer Solar System would provide insight into the environment in which these aggregates were formed. This would provide a unique and direct measurement of Solar System disk properties at the earliest stages of planetary formation.

This material is based upon work supported by the National Science Foundation under Grant No. 0847127. Furthermore, authors are appreciative of the useful discussions with many of the CASPER members.